\documentclass[iop]{emulateapj}
\usepackage{graphicx}

\begin{document}

\title{Evolution of white dwarf stars with high-metallicity progenitors: 
the role of $^{22}$Ne diffusion.}

\author{L. G. Althaus$^{1,2,5}$,
        E. Garc\'{\i}a--Berro$^{2,3}$,
        I. Renedo$^{2,3}$, 
        J. Isern$^{3,4}$
        A. H. C\'orsico$^{1,5}$,
        R. D. Rohrmann$^{6}$}

\affil{$^1$Facultad de Ciencias Astron\'omicas y Geof\'{\i}sicas, 
           Universidad Nacional de La Plata, 
           Paseo del Bosque s/n, 
           (1900) La Plata, 
           Argentina\\
       $^2$Departament de F\'\i sica Aplicada,
           Universitat Polit\'ecnica de Catalunya,
           c/Esteve Terrades 5, 
           08860 Castelldefels,
           Spain\\
       $^3$Institut d'Estudis Espacials de Catalunya,
           c/Gran Capit\`a, 2-4, Edif. Nexus 104, 
           08034 Barcelona, 
           Spain\\
       $^4$Institut de Ci\`encies de l'Espai (CSIC),
           Campus UAB, Facultat de Ci\`encies, Torre C-5, 
           08193 Bellaterra, 
           Spain\\ 
       $^5$Instituto de Astrof\'{\i}sica de La Plata, 
           IALP (CCT La Plata), 
           CONICET-UNLP\\
       $^6$Instituto de Ciencias Astron\'omicas, de la Tierra y del Espacio, CONICET, 
           Av. de Espa\~na 1512 (Sur), CC 49 
           (5500) San Juan, 
           Argentina}
 
\email{althaus@fcaglp.unlp.edu.ar}

\begin{abstract}
Motivated by the strong discrepancy between the main sequence turn-off
age and the white dwarf cooling age in the metal-rich open cluster NGC
6791, we  compute a  grid of white  dwarf evolutionary  sequences that
incorporates for the  first time the energy released  by the processes
of $^{22}$Ne sedimentation and  of carbon/oxygen phase separation upon
crystallization.  The grid covers the  mass range from 0.52 to $1.0 \,
M_{\sun}$,  and it is  appropriate for  the study  of white  dwarfs in
metal-rich  clusters.  The  evolutionary calculations  are based  on a
detailed  and self-consistent  treatment of  the energy  released from
these  two  processes, as  well  as  on  the employment  of  realistic
carbon/oxygen profiles, of relevance for an accurate evaluation of the
energy  released  by carbon/oxygen  phase  separation.   We find  that
$^{22}$Ne  sedimentation strongly  delays  the cooling  rate of  white
dwarfs stemming  from progenitors with high  metallicities at moderate
luminosities, whilst carbon/oxygen  phase separation adds considerable
delays at  low luminosities. Cooling  times are sensitive  to possible
uncertainties  in the  actual value  of the  diffusion  coefficient of
$^{22}$Ne. Changing the diffusion coefficient  by a factor of 2, leads
to  maximum  age differences  of  $\approx  8-20\%$  depending on  the
stellar mass.  We find that the magnitude of the delays resulting from
chemical changes in  the core is consistent with the  slow down in the
white dwarf cooling rate that is required to solve the age discrepancy
in NGC 6791.
\end{abstract}


\section{Introduction}

The  evolution of white  dwarf stars  is a  relatively well-understood
process that  can be basically  described as a simple  cooling problem
(Mestel 1952) in which the decrease in the thermal heat content of the
ions  constitutes the  main source  of luminosity.   Because  of this,
white dwarfs  can be  used as independent  reliable cosmics  clocks to
date a  wide variety of  stellar populations. This fact  has attracted
the attention of numerous researchers over the years, who have devoted
large efforts to study in  detail the evolutionary properties of these
stars.  In  particular, it  is important to  realize that  an accurate
determination of the rate at  which white dwarfs cool down constitutes
a fundamental  issue.  Detailed  evolutionary models for  these stars,
based on  increasing degrees  of sophistication of  their constitutive
physics and energy sources, have  proved to be valuable at determining
interesting  properties of  many Galactic  populations,  including the
disk (Winget  et al.  1987; Garc\'{\i}a--Berro et  al.  1988a; Hernanz
et  al. 1994;  Garc\'\i  a--Berro et  al.  1999), the  halo (Isern  et
al. 1998; Torres et al.   2002) and globular and open clusters (Richer
et al.   1997; Von Hippel  \& Gilmore 2000;  Hansen et al.   2002; Von
Hippel et al.  2006, Hansen et  al.  2007; Winget et al.  2009).  This
important  application of  white dwarf  stars has  also  been possible
thanks to a parallel effort  devoted to the empirical determination of
the  observed white dwarf  cooling sequences  in stellar  clusters, as
well as  the determination of  the luminosity function of  field white
dwarfs, which also provides a measure of the white dwarf cooling rate.

Detailed models  of white dwarfs  require a complete treatment  of the
energy   sources  resulting   from  the   core   cyrstallization.   In
particular, the  release of both latent  heat (Van Horn  1968; Lamb \&
Van Horn 1975) and gravitational  energy due to the change in chemical
composition  induced  by  crystallization  (Stevenson  1980;  Garc\'\i
a--Berro et  al. 1988b;  Segretain et al.   1994; Isern et  al.  1997,
2000) affect considerably the cooling of white dwarfs.  In particular,
compositional  separation at crystallization  markedly slows  down the
cooling.  This, in turn, influences, for instance, the position of the
cut-off of  the disk white  dwarf luminosity function (Hernanz  et al.
1994), which  is essential in obtaining  an independent determination
of the age of the Galactic disk.

However,  a  new  observational   determination  of  the  white  dwarf
luminosity  function of the  cluser NGC  6791 by  Bedin et  al. (2008)
casts  serious  doubts on  the  reliability  of  existing white  dwarf
evolutionary models and  their use as accurate clocks.   NGC 6791 is a
very  old  (8 Gyr)  and  very  metal-rich  ([Fe/H] $\sim  +0.4$)  open
cluster, for which  Bedin et al.  (2008) have  uncovered the faint end
of the  white dwarf cooling track, and  have convincingly demonstrated
the existence  of a peak and  a subsequent cut-off in  the white dwarf
luminosity function  at $m_{\rm F606W}  \approx 28.15$.  Additionally,
Bedin et  al.  (2008) have found  that the age of  the cluster derived
from  the main  sequence  turn-off  technique (8  Gyr)  is in  serious
conflict  with the  age derived  from the  termination of  the cooling
sequence (6  Gyr).  This discrepancy  has strong implications  for the
theory of white dwarf evolution, and  points out at a missing piece of
physics in  the conventional modeling of white  dwarfs. In particular,
Bedin et al.  (2008) have concluded  that the white dwarfs in NGC 6791
have to cool markedly more  slowly than believed in order to reproduce
the  faint peak  and  measured  cut-off in  the  observed white  dwarf
luminosity  function  at the  age  of  the  cluster derived  from  the
well-established main-sequence turn-off technique.

In  view of the  high metallicity  characterizing NGC  6791 ($Z\approx
0.04$), a viable  physical process that can decrease  the cooling rate
of  white dwarfs  appreciably is  the slow  gravitational  settling of
$^{22}$Ne  in  the  liquid  phase.   $^{22}$Ne is  the  most  abundant
impurity  expected in  the  carbon-oxygen interiors  of typical  white
dwarfs.   Its abundance  by mass  reaches $X_{\rm  Ne}  \approx Z_{\rm
  CNO}$, and it is the result  of helium burning on $^{14}$N --- built
up during the CNO cycle of  hydrogen burning.  As first noted by Bravo
et  al.  (1992),  the  two  extra neutrons  present  in the  $^{22}$Ne
nucleus (relative  to $A_i=2Z_i$ nuclei,  being $A_i$ the  atomic mass
number and $Z_i$  the charge) results in a  net downward gravitational
force  of magnitude  $2 m_pg$,  where $g$  is the  local gravitational
acceleration  and $m_p$  is the  proton mass.  This leads  to  a slow,
diffusive  settling of  $^{22}$Ne in  the liquid  regions  towards the
center of the white dwarf.  The role of $^{22}$Ne sedimentation in the
energetics of crystallizing white  dwarfs was first addressed by Isern
et al.  (1991) and more  recently quantitatively explored by Deloye \&
Bildsten (2002)  and Garc\'{\i}a--Berro et al.   (2008), who concluded
that  $^{22}$Ne  sedimentation releases  sufficient  energy to  affect
appreciably the  cooling of massive  white dwarfs, making  them appear
bright  for very  long periods  of time,  of the  order of  $10^9$ yr.
Deloye  \&  Bildsten (2002)  predicted  that  the  possible impact  of
$^{22}$Ne sedimentation on white dwarf cooling could be better seen in
metal-rich  clusters,  such as  NGC  6791,  where  the neon  abundance
expected in the  cores of white dwarfs could be as  high as $\sim 4\%$
by mass.

The effect  of $^{22}$Ne sedimentation is  not included in  any of the
existing  grids  of white  dwarf  evolutionary  calculations, and  its
effect  on the  evolution of  white dwarfs  resulting  from supersolar
metallicity progenitors has not been  addressed. The only study of the
effects  of $^{22}$Ne  sedimentation in  the cooling  of  white dwarfs
using   a   complete   stellar    evolutionary   code   is   that   of
Garc\'{\i}a--Berro et al. (2008) for the case of solar metallicity. In
this paper, we present the first grid of full white dwarf evolutionary
models resulting from metal-rich  progenitors with masses ranging from
1 to  $5 \, M_{\sun}$  that includes both $^{22}$Ne  sedimentation and
carbon-oxygen phase  separation.  This  grid incorporates a  much more
elaborated and  improved treatment of the  physical processes relevant
for   the  white   dwarf  evolution   than  that   we   considered  in
Garc\'{\i}a--Berro  et al.   (2008).  These  improvements  include, in
addition to an  update in the microphysics content,  the derivation of
starting white  dwarf configurations obtained from  a full calculation
of the progenitor evolution, as  well as a precise and self-consistent
treatment of the  energy released by the redistribution  of carbon and
oxygen  due  to  phase  separation during  cystallization,  which  was
lacking in  our previous study.  We  find that the  energy released by
$^{22}$Ne sedimentation markedly impacts the evolution of white dwarfs
populating metal-rich clusters, and that this source of energy must be
taken  into account  in deriving  stellar  ages from  the white  dwarf
cooling sequence of such clusters.  In particular, at the evolutionary
stages where the faint peak  and cut-off of the white dwarf luminosity
function of NGC 6791 are observed,  we find that the release of energy
from both  phase separation and  $^{22}$Ne sedimentation substantially
slows  down the  cooling of  white  dwarfs.  The  occurrence of  these
physical separation processes in the core of cool white dwarfs and the
associated  slow   down  of  the   cooling  rate  has   recently  been
demonstrated by  Garc\'\i a--Berro et  al. (2010) to be  a fundamental
aspect to reconcile the age discrepancy in NGC 6791.

In this  study there are three distinctive  characteristics that allow
us to  obtain absolute ages  for white dwarfs in  metal-rich clusters.
First, as already mentioned, the inclusion of the energy released from
both  $^{22}$Ne sedimentation  and carbon-oxygen  phase  separation is
done  self-consistently  and  locally  coupled  to  the  full  set  of
equations of stellar  evolution.  In addition, realistic carbon-oxygen
profiles expected  in the cores of  white dwarfs, of  relevance for an
accurate evaluation  of the energy  released by phase  separation, are
derived  form the  full  computation of  the  evolution of  progenitor
stars.   Finally,  detailed non-gray  model  atmospheres  are used  to
derive the  outer boundary conditions of our  evolving sequences.  All
these facts allow us to  obtain accurate ages.  The paper is organized
as follows. In Sect.  2 we give a full account of the input physics of
our  evolutionary  code,  particularly  the treatment  of  the  energy
sources.  In Sect.  3 we present  our results, and finally in Sect.  4
we summarize our findings and we draw our conclusions.

\section{Details of computations}

\subsection{Input physics}

Evolutionary calculations for both the white dwarfs and the progenitor
stars were  done with an updated  version of the  {\tt LPCODE} stellar
evolutionary  code ---  see  Althaus et  al.   (2005a) and  references
therein.   This code  has recently  been employed  to  study different
aspects of the evolution of  low-mass stars, such as the formation and
evolution of H-deficient white  dwarfs, PG 1159 and extreme horizontal
branch stars (Althaus et al.  2005a; Miller Bertolami \& Althaus 2006;
Miller Bertolami et al.  2008; Althaus  et al.  2009a), as well as the
evolution of  He-core white  dwarfs with high  metallicity progenitors
(Althaus  et  al.   2009b).   It  has  also been  used  to  study  the
initial-final-mass relation in Salaris et al. (2009), where a test and
comparison of {\tt LPCODE} with other evolutionary codes has also been
made.  Details  of {\tt LPCODE} can  be found in these  works. In what
follows, we  comment on the main  input physics that  are relevant for
the evolutionary calculations presented in this work.

The {\tt LPCODE} evolutionary  code considers a simultaneous treatment
of non-instantaneous  mixing (or extra-mixing if  present) and burning
of  elements (Althaus  et al.   2003).  The  nuclear  network accounts
explicitly for  16 chemical  elements, and the  thermonuclear reaction
rates  are  those described  in  Althaus  et  al.  (2005a),  with  the
exception  of  $^{12}$C$\  +\  $p$  \rightarrow \  ^{13}$N  +  $\gamma
\rightarrow  \   ^{13}$C  +  e$^+  +  \nu_{\rm   e}$  and  $^{13}$C(p,
$\gamma)^{14}$N,  which are  taken from  Angulo et  al.   (1999).  The
$^{12}$C($\alpha,\gamma)^{16}$O reaction rate  is taken from Angulo et
al.  (1999) as  well, and is about twice as large  as that of Caughlan
\& Fowler  (1988).  The final carbon-oxygen composition  is a relevant
issue,  as  a proper  computation  of  the  energy released  by  phase
separation markedly depends on the  chemical profile of the core.  The
standard  mixing  length  theory  for  convection ---  with  the  free
parameter $\alpha=1.61$  --- has been  adopted.  With this  value, the
present luminosity and effective temperature  of the Sun, at an age of
4570 Myr, are reproduced by {\tt LPCODE} when $Z=0.0164$ and $X=0.714$
are  adopted ---  in agreement  with the  $Z/X$ value  of  Grevesse \&
Sauval (1998).

Except   for   the    evolutionary   stages   corresponding   to   the
thermally-pulsing asymptotic  giant branch (AGB)  phase, we considered
the  occurrence  of   extra-mixing  episodes  beyond  each  convective
boundary  following  the  prescription   of  Herwig  et  al.   (1997).
Extra-mixing episodes largely determine  the final chemical profile of
white  dwarfs.  We  treated  extra-mixing as  a  diffusion process  by
assuming  that  mixing  velocities  decay  exponentially  beyond  each
convective boundary.  Specifically, we assumed a diffusion coefficient
given by $D_{\rm OV}= D_{\rm O}\ \exp(-2z/f H_{\rm p})$, where $H_{\rm
  P}$ is the pressure scale height at the convective boundary, $D_{\rm
  O}$ is  the diffusion coefficient  of unstable regions close  to the
convective boundary, and  $z$ is the geometric distance  from the edge
of  the  convective  boundary  (Herwig  et  al.   1997).   We  adopted
$f=0.016$ in all of our sequences,  a value inferred from the width of
the upper main sequence.

Other  physical  ingredients  considered   in  {\tt  LPCODE}  are  the
radiative opacities  from the OPAL project (Iglesias  \& Rogers 1996),
including C- and O-rich  composition, supplemented at low temperatures
with the molecular opacities  of Alexander \& Ferguson (1994).  During
the white dwarf regime, the metal mass fraction $Z$ in the envelope is
not  assumed  to be  fixed.   Instead,  it  is specified  consistently
according  to the  prediction of  element diffusion.   To  account for
this,  we have  considered radiative  opacities tables  from  OPAL for
arbitrary metallicities.  For  effective temperatures less than 10,000
K we have included the  effects of molecular opacitiy by assuming pure
hydrogen  composition  from  the  computations of  Marigo  \&  Aringer
(2009).  This assumption is  justified because element diffusion leads
to pure hydrogen  envelopes in cool white dwarfs.   It is worth noting
that these  opacity calculations do not cover  the high-density regime
characteristic of  the envelopes  of cool white  dwarfs. Nevertheless,
because  the  derivation of  the  outer  boundary  conditions for  our
evolving models  involves the  integration of detailed  non-gray model
atmospheres  down  to  very  large optical  depths  ($\tau=25$)  these
opacities  are  only  required  at  large  $\tau$  and  low  effective
temperatures. However, at the high densities reached at the end of the
atmospheric  integration,  energy transfer  is  mainly by  convection,
which at such depths is essentially adiabatic. Indeed, we find that at
$\tau=25$, the radiative flux  amounts to $4\%$ at most. Consequently,
the  temperature  stratification   characterizing  these  deep  layers
becomes  strongly  tied  to  the  equation of  state,  so  a  detailed
knowledge  of the  radiative opacity  becomes almost  irrelevant.  The
conductive  opacities are  those of  Cassisi et  al.  (2007),  and the
neutrino emission rates are taken from Itoh et al.  (1996) and Haft et
al.   (1994).  For the  high density  regime characteristics  of white
dwarfs,  we  have used  the  equation of  state  of  Segretain et  al.
(1994), which  accounts for all  the important contributions  for both
the  liquid  and solid  phases  --- see  Althaus  et  al.  (2007)  and
references  therein.  We  have also  considered the  abundance changes
resulting from element diffusion in  the outer layers of white dwarfs.
As  a  result, our  sequences  develop  pure  hydrogen envelopes,  the
thickness of which gradually increases as evolution proceeds.  We have
considered gravitational settling  and thermal and chemical diffusion,
see Althaus  et al.  (2003)  for details.  In {\tt  LPCODE}, diffusion
becomes operative  once the  wind limit is  reached at  high effective
temperatures (Unglaub \& Bues 2000).  Chemical rehomogenization of the
inner carbon-oxygen  profile induced by  Rayleigh-Taylor instabilities
has  been   considered  following  Salaris  et   al.   (1997).   These
instabilities  arise because the  positive molecular  weight gradients
that remain above the flat  chemical profile left by convection during
helium core burning.
 
Finally, we  employ outer boundary  conditions for our  evolving white
dwarf models  as provided by detailed non-gray  model atmospheres that
include non-ideal  effects in the  gas equation of state  and chemical
equilibrium based on the  occupation probability formalism.  The level
occupation  probabilities are  self-consistently  incorporated in  the
calculation of  the line  and continuum opacities.   Model atmospheres
also  consider   collision-induced  absorption  due   to  H$_2$-H$_2$,
H$_2$-He, and  H-He pairs, and the  Ly$\alpha$ quasi-molecular opacity
that result from perturbations  of hydrogen atoms by interactions with
other  particles, mainly H  and H$_2$.   These model  atmospheres have
been  developed by Rohrmann  et al.   (2002, 2010),  and we  refer the
reader  to  those  works and  to  Renedo  et  al.  (2010) for  a  full
description of them.   In the interest of reducing  computing time, we
have  computed from  these  models a  grid  of pressure,  temperature,
radial thickness  and outer mass  fraction values at an  optical depth
$\tau=25$  from which  we derive  the outer  boundary  conditions.  At
advanced  stages of  white  dwarf evolution,  the central  temperature
becomes strongly  tied to the  temperature stratification of  the very
outer  layers, thus the  employment of  non-gray model  atmospheres is
highly desired  for an  accurate assessment of  cooling times  of cool
white dwarfs (Prada Moroni  \& Straniero 2007).  Our model atmoshperes
also provide detailed colors and magnitudes for effective temperatures
lower than 60,000K for a pure hydrogen composition and for the HST ACS
filters (Vega-mag system) and $UBVRI$ photometry.

\subsubsection{Energy released from  $^{22}$Ne sedimentation 
and crystallization}

The energy  contribution resulting from the  gravitational settling of
$^{22}$Ne  is   treated  in   a  similar  way   as  it  was   done  in
Garc\'{\i}a--Berro et al. (2008), except that now we have assumed that
the  liquid  behaves  as  a  single  backgroung  one-component  plasma
consisting of the average by number of carbon and oxygen --- the inner
chemical composition expected in a real white dwarf --- plus traces of
$^{22}$Ne. This allows us to  treat the problem of $^{22}$Ne diffusion
in  a simple  and realistic  way.  The  slow change  in  the $^{22}$Ne
chemical  profile  and  the   associated  local  contribution  to  the
luminosity  equation   is  provided   by  an  accurate   treatment  of
time-dependent $^{22}$Ne  diffusion --- see  Garc\'{\i}a--Berro et al.
(2008) for details.  In  particular, the diffusion coefficient $D_{\rm
  s}$ in the liquid interior is given by (Deloye \& Bildsten 2002):

\begin{eqnarray}
D_{\rm s} = 7.3 \times 10^{-7} {{ T } \over{\rho^{1/2} \overline{Z} \ 
\Gamma ^{1/3}}} \ {\rm cm}^2/{\rm s}, 
\label{difucoe}
\end{eqnarray}

\noindent where we have considered a mean charge $\overline{Z}$ of the
background plasma.   For those  regions of the  white dwarf  that have
crystallized, diffusion is  expected to be no longer  efficient due to
the abrupt increase  in viscosity expected in the  solid phase.  Thus,
we  set  $D= 0$  in  the  crystallized  regions.  

In  a subsequent  phase we  have  also considered  the energy  sources
resulting from the crystallization of  the white dwarf core, i.e., the
release  of  latent  heat  and  the release  of  gravitational  energy
associated   with   carbon-oxygen    phase   separation   induced   by
crystallization. In  {\tt LPCODE},  these energy sources  are included
self-consistently and are locally coupled to the full set of equations
of stellar evolution.  In particular, the standard luminosity equation

\begin{equation}
\frac{\partial L_r}{\partial M_r}= \varepsilon_{\rm nuc} -\epsilon_\nu - C_P 
{{d T} \over {d t}} + {{\delta} \over {\rho}}{{d P} \over {d t}} ,
\label{lumistandard}
\end{equation}

\noindent   had  to  be   modified.   In   Eq.   (\ref{lumistandard}),
$\varepsilon_{\rm nuc}$  and $\epsilon_\nu$ denote,  respectively, the
energy per  unit mass per second  due to nuclear  burning and neutrino
losses,   and  the  third   and  fourth   terms  are   the  well-known
contributions of the  heat capacity and pressure changes  to the local
luminosity  of  the  star  (Kippenhahn  \&  Weigert  1990).   We  have
simplified the treatment of phase separation, by ignoring the presence
of $^{22}$Ne.  As shown  by Segretain (1996), $^{22}$Ne influences the
phase diagram at the late stages of crystallization, and the impact on
the cooling time is moderate and much smaller than that resulting from
carbon-oxygen phase separation.  Thus, to compute the energy resulting
from phase separation, we assume that the white dwarf interior is made
only  of carbon  and oxygen  with abundance  by mass  $X_{\rm  C}$ and
$X_{\rm O}$ respectively  ($X_{\rm C} + X_{\rm O}=  1$).  Then, it can
be shown (Garc\'ia--Berro et al. 2008) that

\begin{eqnarray}
\frac{\partial L_r}{\partial M_r}= \varepsilon_{\rm nuc} - \epsilon_\nu - 
C_P{{d T} \over {d t}} + {{\delta} \over {\rho}}{{d P} \over {d t}} + 
&& l_{\rm s}\frac{{dM}_{\rm s}}{dt}\delta(m-M_{\rm s})\nonumber\\
&& -\  A {{d X_{\rm O}} \over {dt} }
\label{lumifin}
\end{eqnarray}

\noindent where $A$ is given by

\begin{equation}
A= \left(\frac{\partial u}{\partial X_{\rm O}}\right)_{\rho,T}+\ {{\delta}\over
{\rho}} \left(\frac{\partial P}{\partial X_{\rm O}} \right)_{\rho,T} \approx 
\left(\frac{\partial u}{\partial X_{\rm O}} \right)_{\rho,T} ,
\label{terminoAfinal}
\end{equation}

\noindent being $u$ the internal energy per gram. 

The fifth  term in  Eq. (\ref{lumifin}) is  the local  contribution of
latent heat:  $l_{\rm s}$  is the latent  heat of  crystallization and
$dM_{\rm s}/dt$ is  the rate at which the solid  core grows. The delta
function  indicates   that  the  latent   heat  is  released   at  the
solidification front.  The last term in Eq. (\ref{lumifin}) represents
the energy released by  chemical abundance changes. Although this term
is usually  small in normal stars,  since it is much  smaller than the
energy  released  by nuclear  reactions,  it  plays  a major  role  in
crystallizing white dwarfs,  with important energetic consequences due
to  carbon-oxygen  phase separation.   In  the  case of  carbon-oxygen
mixtures, $\left(\partial u /  \partial X_{\rm O} \right)_{\rho,T}$ is
domintated by  the ionic contributions,  and is negative.   Hence, the
last term in Eq. (\ref{lumifin}) will  be a source (sink) of energy in
those  regions  where  the  oxygen  abundance  increases  (decreases).
During the crystallization of  a carbon-oxygen white dwarf, the oxygen
abundance  in the  crystallizing region  increases, and  the overlying
liquid mantle becomes carbon-enriched as  a result of a mixing process
induced  by a  Rayleigh-Taylor  instability at  the  region above  the
crystallized  core.  Thus,  according to  Eq.   (\ref{lumifin}), phase
separation will  lead to a source  of energy in those  layers that are
crystallizing, and  to a sink of  energy in the  overlying layers.  We
computed  the   resulting  chemical  rehomogenization   following  the
prescription  by  Salaris  et  al.   (1997), see  also  Montgomery  et
al. (1999).

To implement the  energy release by phase separation in  our code in a
suitable   formalism,  and  to   avoid  numerical   difficulties  when
integrating  the  full  set  of  equations of  stellar  structure  and
evolution, we  have considered  the {\sl net}  energy released  by the
process of  carbon-oxygen phase separation over a  time interval $dt$,
by integrating  the last term  in Eq.  (\ref{lumifin}) over  the whole
star. Because cooling  is a slow process, it can  be shown that (Isern
et al. 1997):

\begin{eqnarray}
\int^{M}_0 \left(\frac{\partial u}{\partial X_{\rm O}}\right)
&&\frac{dX_{\rm O}}{dt}\,d M_r=
(X_{\rm O}^{\rm sol}-X_{\rm O}^{\rm liq})\, \nonumber\\
&& \times\ \left[\left(\frac{\partial u}
{\partial X_{\rm O}}\right)_{M_{\rm s}}-\left\langle\frac{\partial u}
{\partial X_{\rm O}}\right\rangle\right]\frac{dM_{\rm s}}{dt}
\label{total}
\end{eqnarray}

\noindent where $\left(\partial u / \partial X_{\rm O} \right)_{M_{\rm
s}}$ is evaluated at the boundary of the solid core and

\begin{eqnarray}
\left\langle\frac{\partial u}{\partial X_{\rm O}}\right\rangle=\frac{1}{\Delta
M}\int_{\Delta M}\left(\frac{\partial u}{\partial X_{\rm O}}\right) d M_r.
\label{prom}
\end{eqnarray}

The first term  in the square bracket in  Eq. (\ref{total}) represents
the energy released  in the crystallizing layer, and  the second term,
given by  Eq. (\ref{prom}), is the  energy absorbed on  average in the
convective   region  ($\Delta  M$)   driven  by   the  Rayleigh-Taylor
instability  above   the  cyrstallization  front.    Since  $(\partial
u/\partial  X_{\rm O})$  is negative  and essentially  depends  on the
density (which  decreases outwards),  the square bracket  is negative,
and thus the  process of phase separation results in  a net release of
energy during  the time  interval $dt$.  It  is clear that  the energy
released by this  process will depend on the  intial oxygen profile at
the  beginning  of the  white  dwarf  phase,  resulting in  a  smaller
contribution in  the case of initially higher  oxygen abundances. Note
that the  shape of  the initial chemical  profile may also  affect the
degree of  mixing in  the liquid layers  and thus the  energy absorbed
there, hence altering the net energy released by the process.

For  computational   purposes,  we   proceed  as  follows.    At  each
evolutionary time step, we  compute the change of chemical composition
resulting from  carbon-oxygen phase separation  using the spindle-type
phase  diagram for a  carbon-oxygen mixture  of Segretain  \& Chabrier
(1993).  Then,  we evaluate  the net energy  released by  this process
during the time step from  Eq. (\ref{total}).  This energy is added to
the, usually smaller, latent heat  contribution, of the order of $0.77
k_{\rm B}T$ per ion.  The resulting energy is distributed over a small
mass range  around the crystallization front, and  the resulting local
contribution    is   added   to    the   luminosity    equation,   Eq.
\ref{lumistandard}.   Finally,  we  also  add  to  this  equation  the
contribution  from  $^{22}$Ne sedimentation  in  the  same  way as  in
Garc\'{\i}a--Berro   et  al.   (2008).    The  {\tt   LPCODE}  stellar
evolutionary code solves iteratively the full set of equations for the
white  dwarf  evolution  with  the  luminosity  equation  modified  as
previously explained.   We mention that the magnitude  of these energy
sources is calculated at each  iteration during the convergence of the
model.   In our  calculations, crystallization  sets in  when  the ion
coupling constant reaches $\Gamma  =180$, where $\Gamma \equiv \langle
Z^{5/3}\rangle  e^2/a_{\rm  e} k_{\rm  B}T$  and  $a_{\rm  e}$ is  the
interelectronic distance.

Finally,  we  want to  mention  that  our  treatment is  not  entirely
consistent  in the  sense  that the  energy  resulting from  $^{22}$Ne
sedimentation  is  evaluated   separately  and  independently  of  the
$^{22}$Ne abundances  changes induced by  crystallization. However, as
shown  by Segretain  (1996),  the neon  concentration  is expected  to
change  appreciably only  when  $\sim  70\%$ of  the  white dwarf  has
crystallized.   Because  the  luminosity contribution  from  $^{22}$Ne
sedimentation strongly  declines by the  time a large fraction  of the
mass  of  the  white  dwarf  has crystallized,  we  expect  that  this
inconsistency in our treatment is not relevant. In any case, it should
be noted  that this picture could  change appreciably for  the case of
larger  initial  neon abundances  than  that  considered in  Segretain
(1996).

\subsection{Evolutionary sequences}

As we  mentioned, initial  models for our  white dwarf  sequences have
been derived  from full evolutionary calculations  of progenitor stars
for solar  metallicity (Renedo et  al. 2010).  All the  sequences have
been  computed  from  the   ZAMS  through  the  thermally-pulsing  and
mass-loss phases on  the AGB and, finally, to  the domain of planetary
nebulae.   Extra-mixing episodes  beyond  the pulse-driven  convection
zone have been disregarded  during the thermally-pulsing AGB phase, as
suggested  by  different  and   recent  studies  ---  see  Salaris  et
al. (2009),  Weiss \&  Ferguson (2009) and  references therein.   As a
result,  the efficiency  of the  third dredge-up  episode  is strongly
reduced for the low-mass sequences (but not for the massive ones), and
thus the mass of the  hydrogen-free core of our less massive sequences
gradually  grows as evolution  proceeds through  this stage.  A strong
reduction  of extra-mixing  during  the AGB  phase  and the  resulting
reduction in  the efficiency of  third dredge-up episodes  in low-mass
stars  is in  agreement with  observational inferences  of  AGB carbon
stars, the luminosities of which,  in turn, are in good agreement with
those predicted by stellar  models using the Schwarzschild's criterion
for  convection  (Guandalini  et  al.   2006).   The  breathing  pulse
instability  occurring towards  the  end of  core  helium burning  was
suppressed --- see Straniero et  al.  (2003) for a thorough discussion
on this issue.  We considered  mass-loss episodes during the stages of
core helium burning and red giant branch following Schr\"oder \& Cuntz
(2005),  whereas during the  AGB and  thermally-pulsing AGB  phases we
used the mass-loss prescription of Vassiliadis \& Wood (1993).  In the
case  of a  strong reduction  of  the third  dredge-up efficiency,  as
occurs in our less massive sequences,  mass loss plays a major role in
determining the final mass of the hydrogen-free core at the end of the
TP-AGB evolution,  and thus the initial-final mass  relation (Weiss \&
Ferguson  2009).  However,  we  stress  that  the  initial-final  mass
relation obtained from our sequences  (Renedo et al.  2010) is in very
good agreement with the  semi-empirical determination of this relation
of   Salaris  et   al.   (2009)   and  with   that  of   Catal\'an  et
al. (2008). Finally, we mention  that the hydrogen envelope massses of
our sequences should be considered as upper limits to the maximum mass
of  hydrogen left in  a white  dwarf resulting  from the  evolution of
single star progenitors. This stems  from the fact that the occurrence
of a late thermal pulse after departure from the TP-AGB may reduce the
hydrogen mass  considerably, see Althaus et al.  (2005b).  Hence, this
could alter the quantitative  effect of $^{22}$Ne sedimentation on the
white dwarf cooling.

\begin{figure}
\begin{center}
\includegraphics[clip,width=0.9\columnwidth]{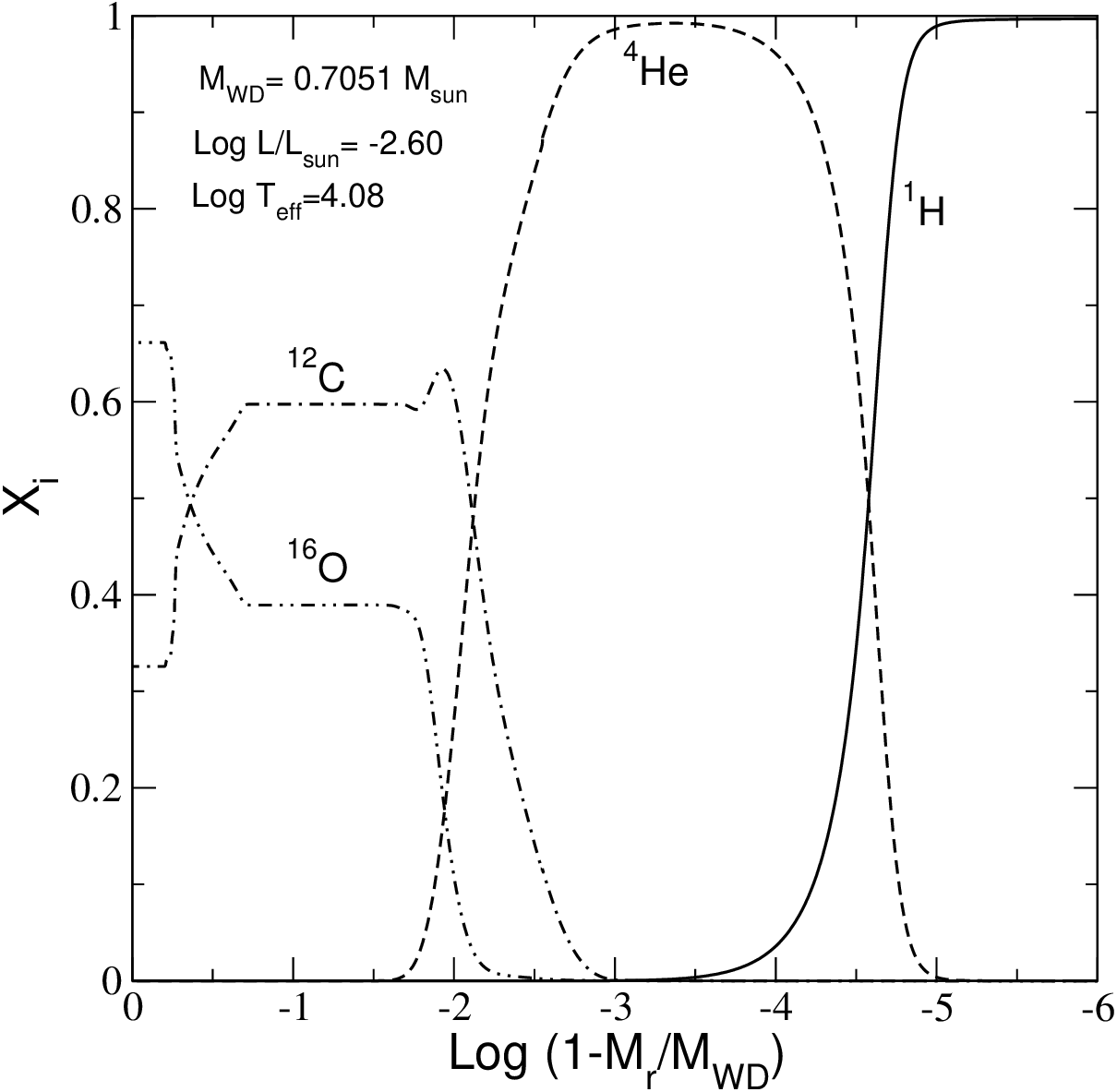}
\caption{The chemical abundance distribution (carbon, oxygen, hydrogen
         and helium)  for a selected $0.7051 \,  M_{\sun}$ white dwarf
         model after element diffusion has  led to the formation of a
         pure hydrogen envelope.}
\label{quimiini}
\end{center}
\end{figure}

The computation of the progenitor evolution provides realistic initial
models   and,  more   importantly,  detailed   carbon-oxygen  chemical
profiles, which  are relevant for  a proper computation of  the energy
released by  carbon-oxygen phase separation.   In Fig.  \ref{quimiini}
we show  the mass abundances  of $^1$H, $^4$He, $^{12}$C  and $^{16}$O
throughout the  deep interior of a selected  $0.7051\, M_{\sun}$ white
dwarf  model at  an  evolutionary stage  where  element diffusion  has
already   strongly   modified  the   initial   outer  layer   chemical
stratification,  leading to  the formation  of a  thick  pure hydrogen
envelope  plus an extended  inner tail.   Below the  hydrogen envelope
there is  the helium buffer and  an intershell rich  in helium, carbon
and  oxygen.  Finally,  the  innermost region  is  composed mainly  of
carbon and oxygen,  plus traces of heavier element  of which $^{22}$Ne
is the most  abundant one.  As previously mentioned,  $^{22}$Ne is the
result  of  helium  burning   on  $^{14}$N  via  the  reactions  \mbox
{$^{14}$N($\alpha,         \gamma$)$^{18}$F($\beta^+$)$^{18}$O($\alpha,
\gamma$)$^{22}$Ne}.  The core chemical profile of our model is typical
of  situations  in  which  extra  mixing  episodes  beyond  the  fully
convective  core  during the  core  He  burning  are allowed  ---  see
Straniero et al. (2003) and  also Prada Moroni \& Straniero (2007) for
the consequences on white  dwarf evolution.  The flat chemical profile
towards  the center  is the  result of  the  chemical rehomogenization
induced by Rayleigh-Taylor instabilities.

\begin{table}
\centering
\scriptsize
\caption{Initial  and final  stellar mass  (in solar  units),  and the
         central  oxygen  abundance (mass  fraction)  of our  starting
         white   dwarf  sequences.   The  progenitor   metallicity  is
         $Z=0.01$.}
\begin{tabular}{llc}
\hline
\hline
$M_{\rm WD}$ &$M_{\rm ZAMS}$  &  $X_{\rm O}$ \\
\hline
 0.5249   & 1.00 & 0.70 \\
 0.5701   & 1.50 & 0.68 \\
 0.5932   & 1.75 & 0.70 \\
 0.6096   & 2.00 & 0.72 \\
 0.6323   & 2.25 & 0.75 \\
 0.6598   & 2.50 & 0.72 \\
 0.7051   & 3.00 & 0.66 \\
 0.7670   & 3.50 & 0.65 \\
 0.8779   & 5.00 & 0.61 \\
\hline
\hline
\end{tabular}
\label{tableini}
\end{table}  

In this work,  we have considered two initial  $^{22}$Ne abundances of
$X_{\rm  Ne}  \approx  Z_{\rm   CNO}\approx  0.03$  and  0.06.   These
elections  are   not  entirely   consistent  with  the   almost  solar
metallicity  we  assumed for  the  progenitor  stars, $Z=0.01$.   This
introduces  a  slight  inconsistency   however,  since  only  a  minor
difference  is expected  in the  oxygen composition  and in  the white
dwarf  evolution  when progenitors  with  different metallicities  are
considered  (Prada Moroni  \& Straniero  2002, Salaris  et  al. 2010).
Thus, to a good  approximation, our starting models are representative
of white  dwarf stars with progenitors  having supersolar metallicity.
However,  we mention  that  a significant  change  in the  metallicity
progenitor could  affect the AGB  evolution and mass-loss  history, as
well as  the initial-final  mass relation. In  this work,  we computed
white  dwarf  sequences  with  hydrogen-rich envelopes  for  $^{22}$Ne
abundances  of   0.03  and  0.06,  taking  into   account  the  energy
contributions  from $^{22}$Ne  sedimentation  and carbon-oxygen  phase
separation.  We compute also additional sequences to assess the impact
of  these  energy  sources.   This  includes the  computation  of  the
evolution  of  a $1.0  \,  M_{\sun}$  white  dwarf sequence  that  was
started,   in    contrast   to   the   other    sequences,   from   an
artificially-generated  initial   model,  and  with   a  carbon-oxygen
composition similar to that of  the $0.8779 \, M_{\sun}$ sequence.  In
this way, our sequences cover the entire white dwarf mass interval for
which  a  carbon-oxygen core  is  expected  to  be formed.   In  Table
\ref{tableini}  we list  the stellar  masses of  the white  dwarfs for
which we compute their  progenitor evolution, together with the inital
masses  of the progenitor  stars at  the ZAMS.   Also listed  in Table
\ref{tableini} is the central oxygen abundance at the beginning of the
white dwarf  evolutionary track.   These sequences were  computed from
the pre-white dwarf stage down to $\log(L/L_{\sun}) \approx -5.3$.  To
explore the  relevance for the  cooling times of uncertainties  in the
actual  value of  the diffusion  coefficient of  $^{22}$Ne  (Deloye \&
Bildsten 2002),  we compute additional cooling  sequences altering the
diffusion coefficient by  a factor of 2.  Finally,  we find worthwhile
to  assess the  lowest metallicity  for which  $^{22}$Ne sedimentation
starts to affect  significantly the cooling times of  white dwarfs. To
this  end,  we  compute   additional  cooling  sequences  for  initial
$^{22}$Ne abundances of 0.01 and 0.005.

\section{Results}

The  results  presented in  this  work are  based  on  a complete  and
consistent treatment  of the  different energy sources  that influence
the  evolution  of  white   dwarfs  along  the  distinct  evolutionary
stages. The  ultimate aim  is to provide  cooling ages as  accurate as
possible, according  to our best  knowledge of the  physical processes
that drive  the evolution of  these stars.  In particular,  we compute
here  the  first  grid  of  white dwarf  evolutionary  sequences  that
incorporates the sedimentation of $^{22}$Ne.  The grid is intended for
applications to  white dwarfs with high $^{22}$Ne  abundances in their
cores, namely, those resulting  from metal-rich progenitors, for which
$^{22}$Ne sedimentation is expected to impact their evolution.  In the
interest  of avoiding  a lengthy  discussion of  the results,  we will
focus on the consequences of $^{22}$Ne sedimentation on the evolution,
postponing  a comprehensive description  of the  standard evolutionary
aspects, particularly the role of carbon-oxygen phase separation, to a
companion publication (Renedo et al. 2010).

As shown  by Deloye \&  Bildsten (2002) and Garc\'{\i}a--Berro  et al.
(2008),  $^{22}$Ne sedimentation is  a slow  process that  impacts the
evolution of white  dwarfs only after long enough  times have elapsed.
During the evolutionary  stages where most of the  white dwarf remains
in a liquid state, this process causes a strong depletion of $^{22}$Ne
in the outer  region of the core, and an  enhancement of its abundance
in  the   central  regions  of   the  star.   This   behavior  becomes
susbtantially more noticeable as  the gravity is increased.  Indeed, a
more rapid  sedimentation and a  faster depletion of $^{22}$Ne  in the
outer layers  is expected in  massive white dwarfs.   However, because
massive  white  dwarfs crystallize  earlier  than  less massive  ones,
$^{22}$Ne sedimentation will stop  at higher effective temperatures as
compared with less  massive white dwarfs, thus limiting  the extent to
which $^{22}$Ne  diffusion constitutes an energy source  for the star.
This is  a critical  issue regarding the  cooling behavior  of massive
white dwarfs.

\begin{figure}
\begin{center}
\includegraphics[clip,width=0.9\columnwidth]{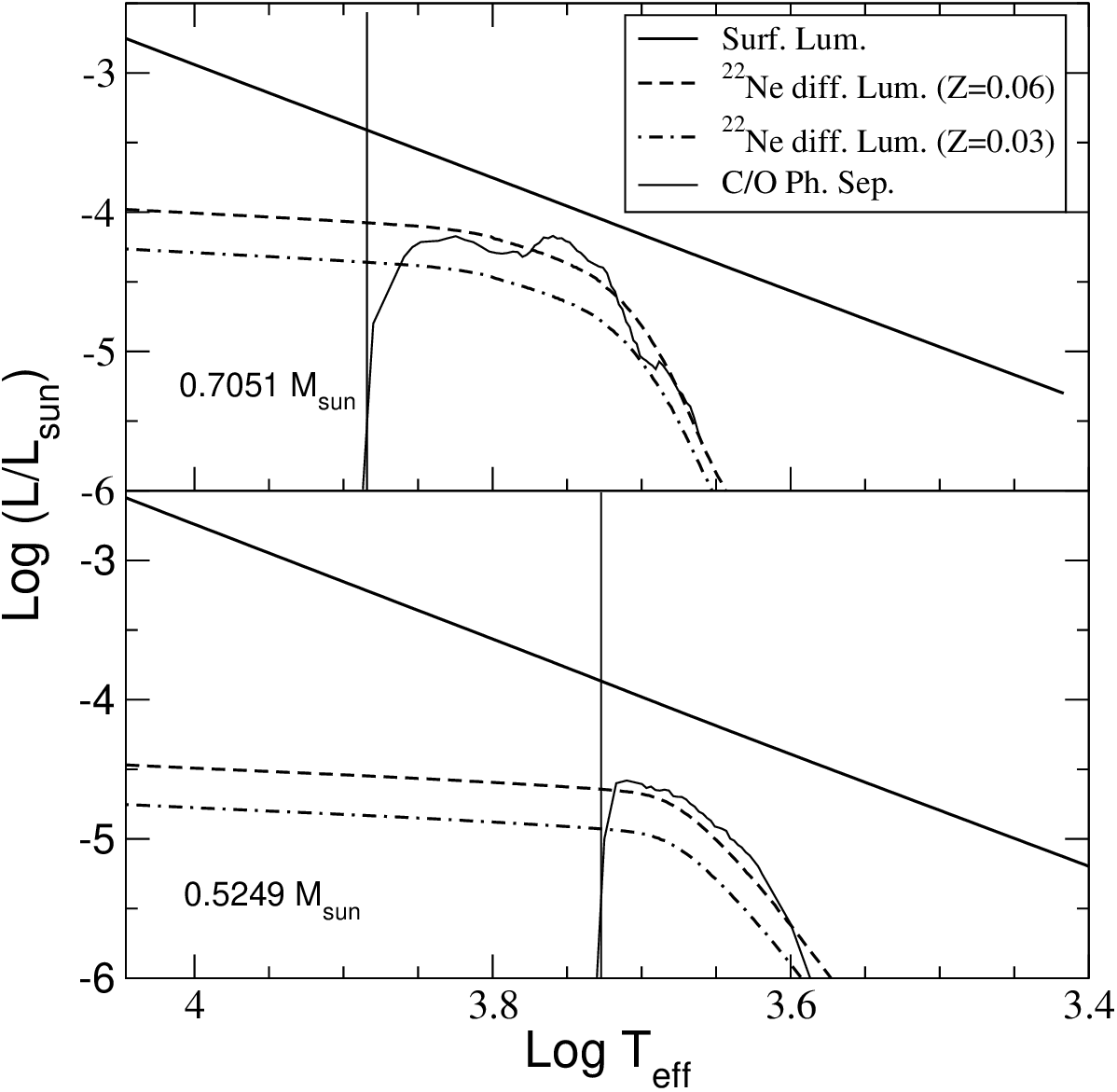}
\caption{Luminosity  contribution  in  solar  units due  to  $^{22}$Ne
         sedimentation versus effective temperature for the 0.7051 and
         $0.5249 \, M_{\sun}$ white  dwarf sequences (upper and bottom
         panel, respectively),  and for $Z=0.06$ and  0.03 (dashed and
         dot-dashed lines) respectively.   The solid line displays the
         surface luminosity,  and the  thin solid line  the luminosity
         contribution   from  carbon-oxygen  phase   separation.   The
         vertical line  marks the effective temperature  for the onset
         of core crystallization.}
\label{hr}
\end{center}
\end{figure}

As  expected,  the  contribution  of $^{22}$Ne  sedimentation  to  the
luminosity budget of white dwarfs  becomes larger as the metal content
of the parent star is increased. This is exemplified in Fig. \ref{hr},
which shows the resulting  luminosity contribution (expressed in solar
units) in  terms of the effective  temperature of the  white dwarf for
the 0.7051  and the $0.5249  \, M_{\sun}$ sequences (upper  and bottom
panel,  respectively) and for  the two  metallicities adopted  in this
work, $Z=0.03$ and  0.06.  This figure gives us a  deep insight of the
importance  of  $^{22}$Ne  sedimentation  into the  global  energetics
during the  entire white dwarf evolution.  Note  that the contribution
from this  process to the star  luminosity is notably  enhanced in the
case of more  massive white dwarfs. Moreover, after  the onset of core
crystallization $^{22}$Ne  sedimentation is still  a relevant process.
As the  core becomes increasingly crystallized, the  luminosity due to
$^{22}$Ne   sedimentation  declines   steeply  (at   higher  effective
temperatures  in  more  massive  white dwarfs).   For  low-mass  white
dwarfs,  the  impact  of  $^{22}$Ne  sedimentation  is  markedly  less
noticeable, albeit  not negligible in the  case of high  $Z$.  In Fig.
\ref{hr},  we  also  display  with  thin solid  lines  the  luminosity
contribution that results from carbon-oxygen phase separation.  It can
been seen that,  depending on the stellar mass  and metal content, the
contribution   of  $^{22}$Ne  sedimentation   to  the   energetics  is
comparable  or larger  than  that resulting  from carbon-oxygen  phase
separation.

\begin{figure}
\begin{center}
\includegraphics[clip,width=0.9\columnwidth]{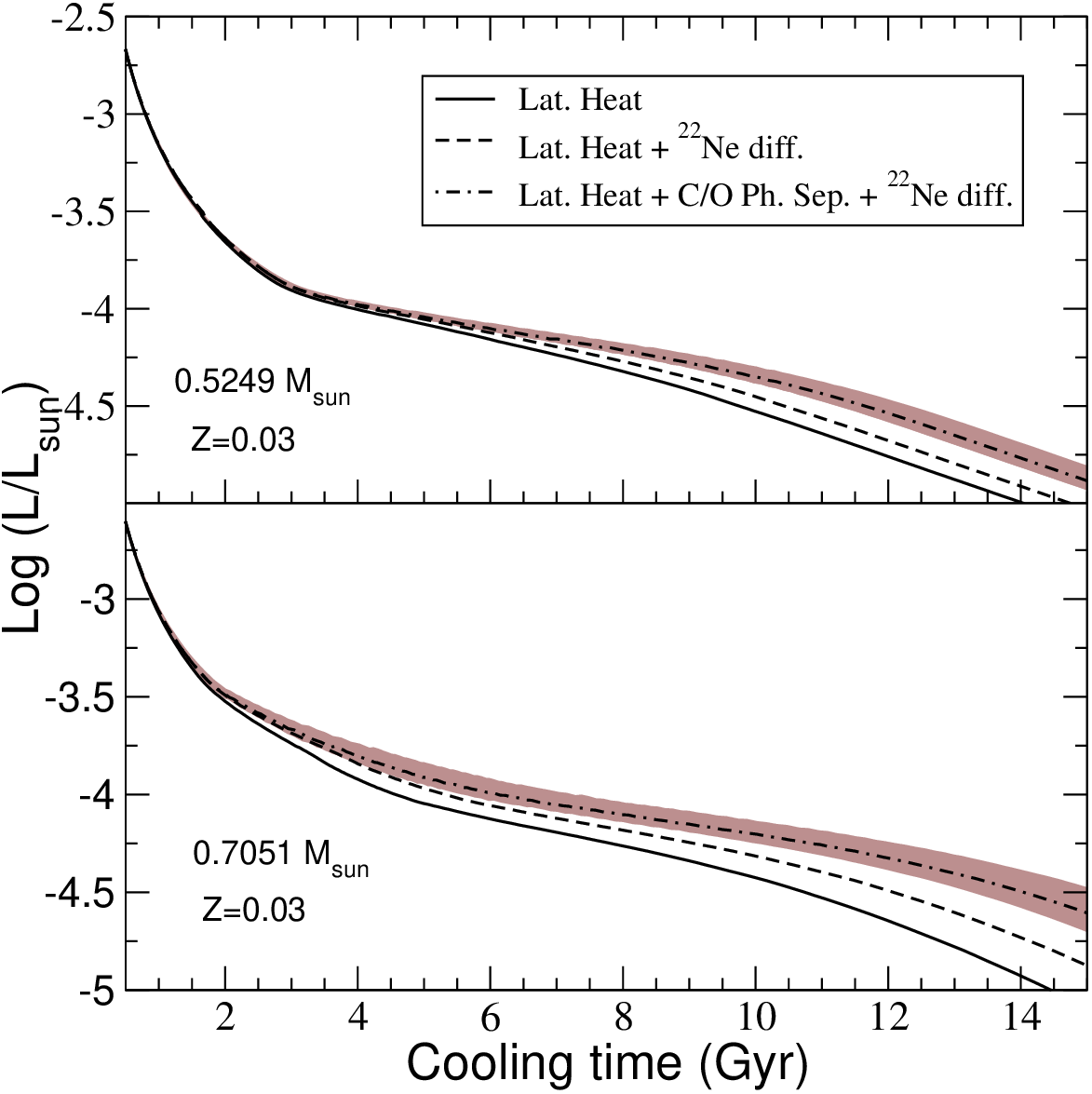}
\caption{Surface luminosity  versus age for the 0.5249  and $0.7051 \,
         M_{\sun}$     sequences    (upper    and     bottom    panel,
         respectively). The solid line  displays the cooling times for
         the  case  in  which  only  the release  of  latent  heat  is
         considered and  carbon-oxygen phase separation  and $^{22}$Ne
         sedimentation  are neglected.  The  dashed line  displays the
         results  for the case  where both  latent heat  and $^{22}$Ne
         sedimentation  are  included,  but  not  carbon-oxygen  phase
         separation. The dot-dashed line corresponds to the case where
         latent  heat, carbon-oxygen  phase  separation and  $^{22}$Ne
         sedimentation are considered. For  this case, the gray region
         shows the extent  to which the cooling times  change when the
         diffusion coefficient of $^{22}$Ne  is changed by a factor of
         2. The metal content in all cases is $Z=0.03$.}
\label{edad_003}
\end{center}
\end{figure}

\begin{figure}
\begin{center}
\includegraphics[clip,width=0.9\columnwidth]{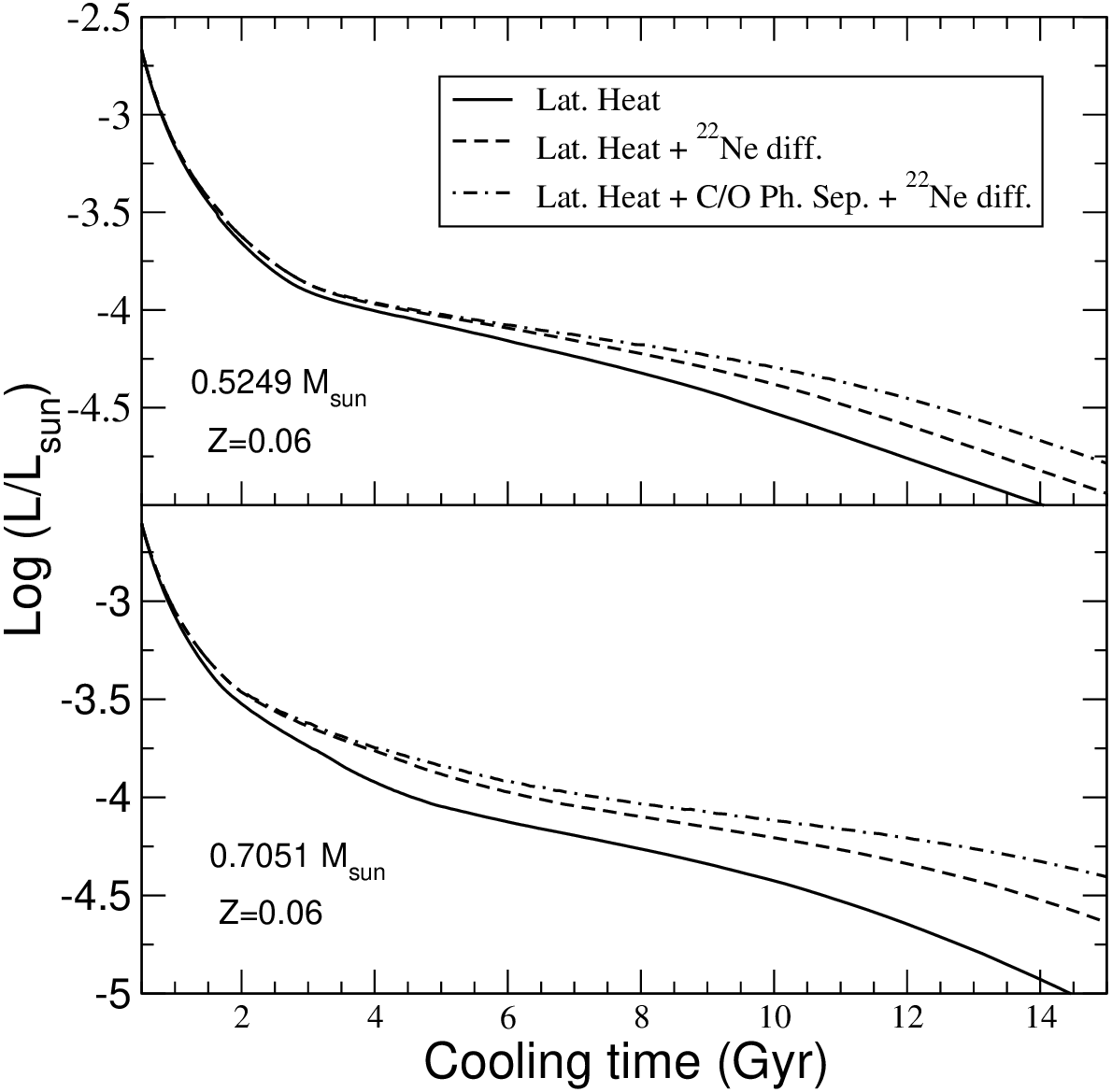}
\caption{Similar to  Fig.  \ref{edad_003}  but for white  dwarf models
         with a metal content of $Z=0.06$ in their cores.}
\label{edad_006}
\end{center}
\end{figure}

It is  clear that  $^{22}$Ne sedimentation plays  a major role  in the
energetics of cool white dwarfs  characterized by a high metal content
in their  interiors. The impact of  this process as well  as of latent
heat  and carbon-oxygen phase  separation in  the white  dwarf cooling
ages can be  seen in Figs.  \ref{edad_003} and  \ref{edad_006} for the
case  of  $Z=0.03$ and  0.06,  respectively.   Here,  the white  dwarf
surface luminosity is shown as a function of the age.  In each figure,
the upper  and bottom  panles correspond to  the 0.5249  and 0.7051$\,
M_{\sun}$ sequences, respectively.  The  solid line corresponds to the
standard case  in which latent  heat is considered,  and carbon-oxygen
phase  separation  and  $^{22}$Ne  sedimentation are  neglected.   The
inclusion  of $^{22}$Ne  sedimentation strongly  modifies  the cooling
curves --- dashed lines. Finally, the addition of the energy resulting
from   carbon-oxygen  phase   separation  upon   cyrstallization  (and
$^{22}$Ne  sedimentation) gives  rise to  the cooling  curve  shown in
dot-dashed   line.   Clearly,   the  energy   released   by  $^{22}$Ne
sedimentation  markedly  influences  the cooling  times,  particularly
those of  the more massive white  dwarfs. Note that in  this case, the
magnitude of the delays in  the cooling rates resulting from $^{22}$Ne
sedimentation  are comparable  (or even  much  larger in  the case  of
$Z=0.06$) to the delays induced by carbon-oxygen phase separation.

\begin{figure}
\begin{center}
\includegraphics[clip,width=0.9\columnwidth]{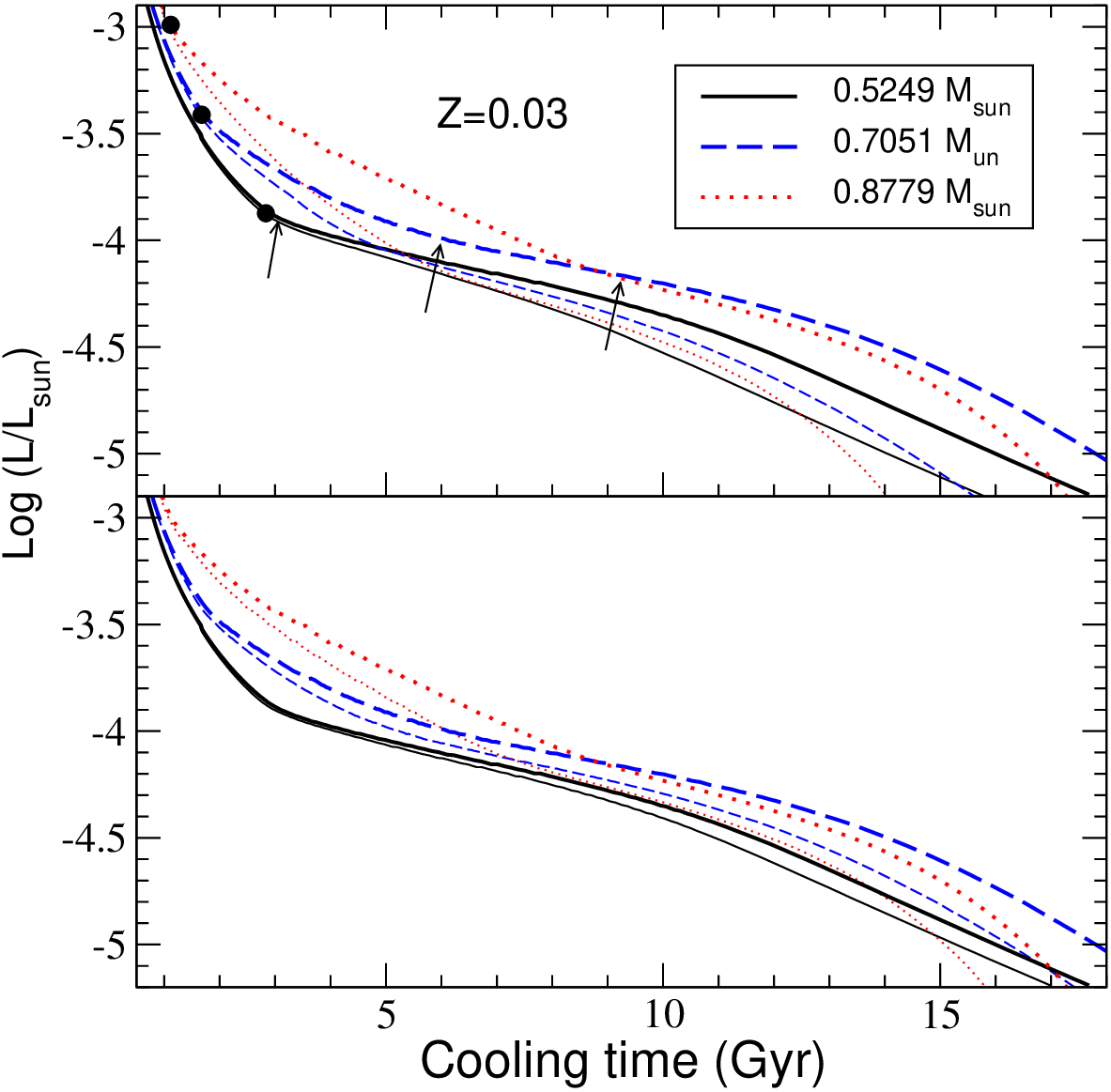}
\caption{Surface  luminosity  versus  cooling  time  for  the  0.5249,
         0.7051, and  $0.8779 \, M_{\sun}$ sequences.  In both panels,
         the thick lines correspond  to sequences where the release of
         latent  heat, carbon-oxygen  phase  separation  and $^{22}$Ne 
         sedimentation  are  considered.   Sequences with  thin  lines
         consider either only latent heat (upper panel) or latent heat
         plus  carbon-oxygen   separation  (botton  panel),   but  not
         $^{22}$Ne  sedimentation.   Filled   circles  and  arrows  at
         selected  sequences   denote,  respectively,  the   onset  of
         crystallization  and the  time at  which  convective coupling
         occurs.  The metallicity is $Z=0.03$.}
\label{masas}
\end{center}
\end{figure}

According  to what  we  have discussed,  the  signatures of  $^{22}$Ne
sedimentation  in  the  cooling   rate  will  certainly  be  different
depending on the  mass of the white dwarf.   In particular, because of
their larger  gravities, they start to manifest  themselves earlier in
more massive  white dwarfs.   This can be  better appreciated  in Fig.
\ref{masas}, where we show the  cooling curves for the 0.5249, 0.7051,
and  $0.8779 \,M_{\sun}$  sequences  for the  case  $Z=0.03$. In  both
panels, we  display with  thick lines the  cooling curves  that result
from considering the energy released by crystallization (latent heat),
carbon-oxygen phase separation  and $^{22}$Ne sedimentation.  The thin
lines   show   the  corresponding   cooling   curves  when   $^{22}$Ne
sedimentation is neglected, but  latent heat (upper panel), and latent
heat and carbon-oxygen phase  separation (bottom panel) are taken into
account.   Note  the marked  lengthening  of  the  cooling times  that
results from the ocurrence  of $^{22}$Ne sedimentation at luminosities
as high as $\log(L/L_{\sun}) \approx  -3.5$ (see bottom panel) for the
massive sequence of $0.8779 \,M_{\sun}$. This delay persists until low
luminosities.  For  the lowest stellar masses considered  in this work
($0.5249 \,M_{\sun}$), appreciable delays  in the cooling rates due to
$^{22}$Ne  sedimentation take  place, but  only at  luminosities lower
than $\log(L/L_{\sun}) \approx  -4.2$.  Carbon-oxygen phase separation
also leads to appreciable delays  in the cooling rates.  The extent of
this delay  and the  luminosities at which  this occurs depend  on the
stellar mass.  In  the case of more massive white  dwarfs, most of the
carbon-oxygen  phase  separation  and  $^{22}$Ne  sedimentation  occur
during evolutionary  stages prior to  the convective coupling  --- the
onset  of which is  indicated by  arrows on  the sequences  with thick
lines in the  upper panel.  Convective coupling takes  places when the
envelope convective  region penetrates into the  degenerate core, with
the  consequent release of  excess thermal  energy, and  the resulting
slow-down of the cooling process,  as reflected by the change of slope
in the cooling curve --- see  Fontaine et al.  (2001). By contrast, in
the  less  massive  models, the  delay  in  the  cooling rate  due  to
convective  coupling takes  place  before the  release of  appreciable
energy   from    carbon-oxygen   phase   separation    and   $^{22}$Ne
sedimentation.  This in part helps to understand the distinct behavior
of the cooling curves with stellar mass.

\begin{figure}
\begin{center}
\includegraphics[clip,width=0.9\columnwidth]{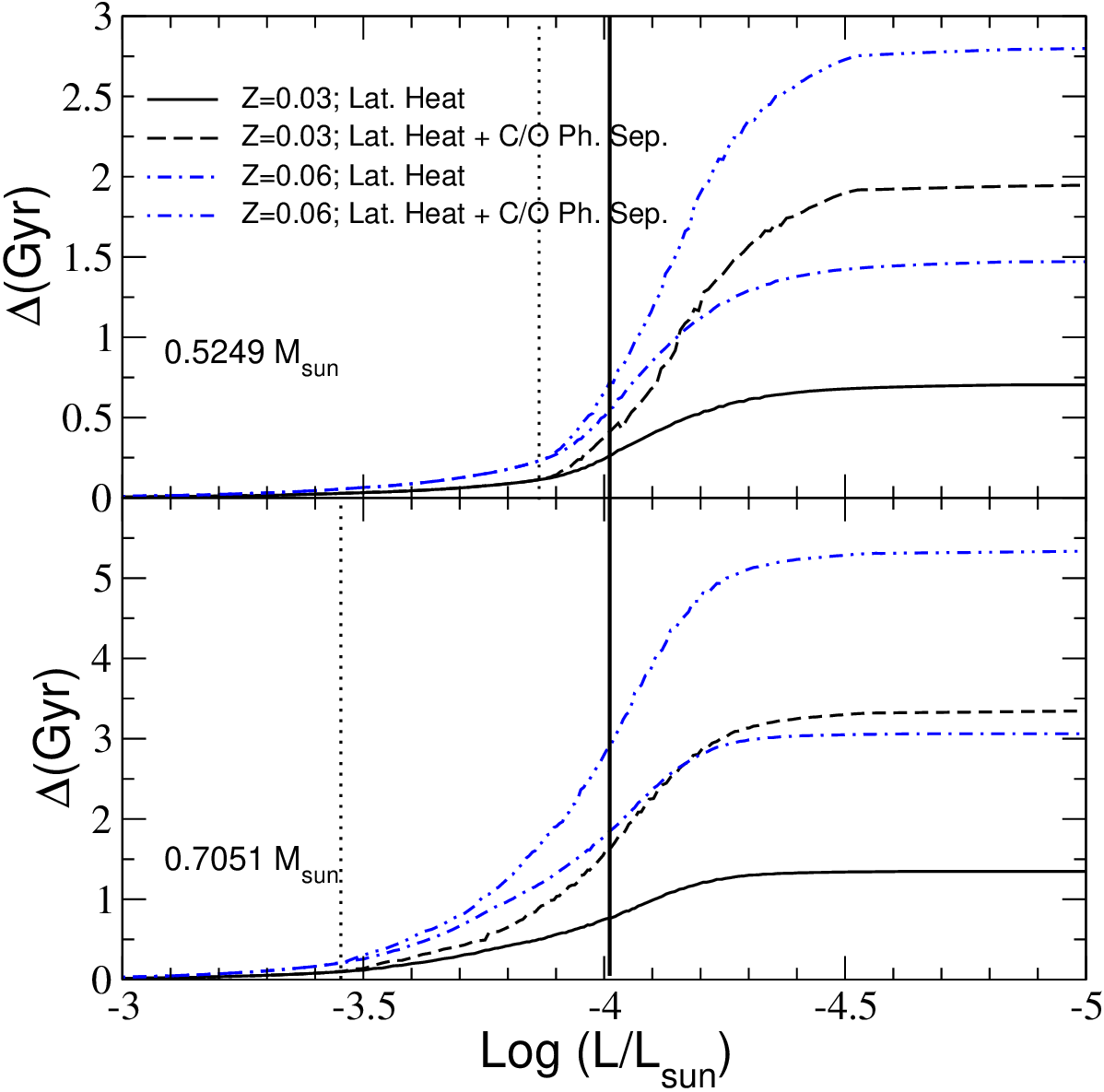}
\caption{Difference  in  evolutionary   times  (Gyr)  between  various
         sequences that  include $^{22}$Ne diffusion  and the sequence
         which  considers  only latent  heat,  for  white dwarfs  with
         masses 0.5249 and $0.7051  \, M_{\sun}$.  The vertical dotted
         lines  mark  the surface  luminosity  at  the  onset of  core
         crystallization, and the vertical  solid line the location of
         the faint peak in the NGC 6791 luminosity function.}
\label{comparo}
\end{center}
\end{figure}

\begin{table*}
\scriptsize
\setlength{\tabcolsep}{0.02in} 
\caption{Differences in the evolutionary times (Gyr) between sequences
         which include the release of latent heat, carbon-oxygen phase
         separation  and  $^{22}$Ne  sedimentation with  the  sequence
         which considers only the release of latent heat.  The results
         are  shown at  selected stellar  lumonisities and  masses (in
         solar units), for both $Z=0.03$ and $Z=0.06$.}
\begin{center}
\begin{tabular}{ccccccccccccccccccc}
\hline
\hline
{\ } 
&\multicolumn{7}{c}{$Z=0.03$} 
&{\ }
&\multicolumn{7}{c}{$Z=0.06$} \\
\cline{2-8} 
\cline{10-16}
$\log(L/L_{\sun})$ & 
$0.5249$ & 
$0.5932$ & 
$0.6598$ & 
$0.7051$ & 
$0.7670$ &
$0.8779$ &
$1.0000$ &
{\ }     & 
$0.5249$ & 
$0.5932$ & 
$0.6598$ & 
$0.7051$ &
$0.7670$ &
$0.8779$ &
$1.0000$\\
\cline{1-8}
\cline{10-16}
 $-3.0$ & $<0.01$ & 0.01 & 0.01 & 0.02 & 0.03 & 0.07 & 0.35 && 0.01 & 0.02 & 0.03 & 0.03 & 0.05 & 0.14 & 0.67\\
 $-3.5$ & 0.03    & 0.05 & 0.10 & 0.14 & 0.33 & 1.02 & 1.45 && 0.07 & 0.09 & 0.20 & 0.30 & 0.64 & 1.80 & 2.61 \\
 $-4.0$ & 0.38    & 0.64 & 1.03 & 1.63 & 2.04 & 2.51 & 2.30 && 0.66 & 1.18 & 1.89 & 2.88 & 3.58 & 4.26 & 3.79 \\
 $-4.5$ & 1.90    & 2.44 & 2.84 & 3.31 & 3.33 & 3.20 & 2.60 && 2.73 & 3.69 & 4.43 & 5.29 & 5.33 & 5.05 & 4.10 \\
\hline
\hline
\end{tabular}
\end {center}
\label{delay}
\end{table*}

From the preceeding paragraphs, we conclude that, in the case of white
dwarfs that result from progenitor stars with super-solar metallicity,
the occurrence  of $^{22}$Ne  sedimentation releases enough  energy to
produce appreciable delays in the  rate of cooling at relevant stellar
luminosities.   We  can  obtain  a  more  quantitative  idea  of  this
assertion examining Fig. \ref{comparo}  --- see also Table \ref{delay}
--- which shows  the age difference between  sequences that incorporte
the energy released from $^{22}$Ne sedimentation and the sequence that
considers only latent  heat, for two masses and  the two metallicities
adopted  in this  study.  It  is clear  that both  carbon-oxygen phase
separation and  $^{22}$Ne sedimentation lead to  substantial delays in
the cooling  times.  Note that,  at very low surface  luminosities and
for metallicity  $Z=0.03$, the  inclusion of both  carbon-oxygen phase
separation  and  $^{22}$Ne   sedimentation  produces  age  differences
between $\sim 2.0$ and 3.3 Gry,  depending on the value of the stellar
mass.  These  differences range from  2.7 to 5.3  Gyr for the  case of
progenitor stars with  $Z=0.06$.  Note also that the  magnitude of the
delay resulting from the $^{22}$Ne sedimentation is comparable or even
larger  than  that  induced  by carbon-oxygen  phase  separation.   In
particular,  by  $\log(L/L_{\sun}) \approx  -4.0$,  the luminosity  at
which the  faint peak  of the white  dwarf luminosity function  in NGC
6791 is  located, the release  of energy from  $^{22}$Ne sedimentation
markedly slows down the cooling rate of the more massive white dwarfs,
which,  because of  their short  pre-white dwarf  times,  populate the
faint end of the white  dwarf luminosity function of the cluster.  For
our  more massive  sequences and  at  the metallicity  of the  cluster
($Z\simeq 0.04$), we find that the delays from $^{22}$Ne sedimentation
alone  range  from  1.10  to  1.50  Gyr  and  $\approx$  1.80  Gyr  at
$\log(L/L_{\sun}) \approx -4.0$ and $-4.2$, respectively. These delays
together   with  the   delays  resulting   from   carbon-oxygen  phase
separation, are  of the  order of  what is required  to solve  the age
discrepancy in NGC 6791 (Garc\'\i a--Berro et al. 2010).  The delay in
the   cooling  rate   of   white  dwarfs   resulting  from   $^{22}$Ne
sedimentation   is  important   and  points   out  the   necessity  of
incorporating this energy source  in the calculation of detailed white
dwarf  cooling sequences,  particularly in  the case  of  white dwarfs
populating metal-rich clusters.

Because  of  the relevance  of  the  $^{22}$Ne  sedimentation for  the
cooling of  white dwarfs, we  find instructive to estimate  the lowest
metallicity  for  which   $^{22}$Ne  sedimentation  starts  to  affect
significantly  the cooling  times of  white  dwarfs. To  this end,  we
compute additional cooling  sequences for initial $^{22}$Ne abundances
of 0.01 and  0.005. For the case of an  initial $^{22}$Ne abundance of
0.01,  $^{22}$Ne  sedimentation  increases  the cooling  time  of  our
$0.5249  \,   M_{\sun}$  sequence  that  considers   latent  heat  and
carbon-oxygen phase  separation by at most 1--2$\%$.  The magnitude of
the delays are larger for more massive white dwarfs, reaching 3--6$\%$
and  4--8$\%$  for  the   0.7051  and  $1.0  \,  M_{\sun}$  sequences,
respectively. For  these two stellar masses, the  resulting delays are
1.5--3$\%$  and  2--4$\%$ for  a  $^{22}$Ne  abundances  of 0.005.  We
conclude that, for initial  $^{22}$Ne abundances smaller than $\approx
0.01$,  $^{22}$Ne sedimentation  has  a minor  impact  on white  dwarf
cooling  times,  except for  rather  massive  white  dwarfs for  which
non-negligible  delays (but  smaller than  $8\%$) are  found  even for
$^{22}$Ne abundances of 0.005.

Finally, to account for possible  uncertainties in the actual value of
the diffusion  coefficient of $^{22}$Ne (Deloye \&  Bildsten 2002), we
compute additional cooling sequences  for which we multiply and divide
the diffusion coefficient  by a factor of 2.  The resulting impacts on
the  cooling time  for  the case  in  which $Z=0.03$  can  be seen  in
Fig.  \ref{edad_003}  for the  0.7051  and  the  $0.5249 \,  M_{\sun}$
sequences  that consider latent  heat, carbon-oxygen  phase separation
and  $^{22}$Ne sedimentation.   The gray  region shows  the  extent to
which  the  cooling curves  vary  when  the  diffusion coefficient  is
changed within this range of values. For the more massive sequence and
at $\log(L/L_{\sun})  = -4.5$  and $-4$, the  cooling times  change by
less than  $8\%$ and $-5\%$,  and by $17\%$ and  $-8\%$, respectively.
For the less massive sequence, the changes remain below $7\%$.  In the
case of the $1.0 \, M_{\sun}$  sequence, an increase in $D_{\rm s}$ by
a factor  of 2 causes a maximum  age difference of $\approx  20 \%$ in
the luminosity range from $\log(L/L_{\sun}) \approx -3$ to $-3.5$.  It
is clear that uncertainties in the diffusion coefficient larger than a
factor of  2 will affect  the cooling time  considerably, particularly
for our most massive white dwarf sequences.

\section{Conclusions}

The use  of white  dwarfs as reliable  cosmic clocks to  date Galactic
stellar  populations has  been recently  thrown  into doubt  by a  new
observational determination of the  white dwarf luminosity function in
the old,  metal-rich open cluster NGC  6791 (Bedin et  al.  2008), the
age of which  as derived from the main  sequence turn-off technique (8
Gyr) markedly disagrees  with the age derived from  the termination of
the white dwarf cooling sequence (6 Gyr).  This discrepancy points out
at a missing physical process in the standard treatment of white dwarf
evolution.  In  view of the  high metallicity characterizing  NGC 6791
($Z\approx 0.04$), the gravitational settling of $^{22}$Ne constitutes
the most  viable process  that can decrease  the cooling rate  of cool
white dwarfs.  Indeed, as first shown by Isern et al. (1991) and later
by Deloye \& Bildsten (2002) and Garc\'{\i}a--Berro et al. (2008), the
slow gravitational settling of  $^{22}$Ne in the liquid phase releases
enough energy  as to appreciably slow  down the cooling  rate of white
dwarfs in metal-rich clusters like NGC 6791.

Motivated by these  considerations, we have presented a  grid of white
dwarf evolutionary sequences that  incorporates for the first time the
energy  contributions arising  from both  $^{22}$Ne  sedimentation and
carbon-oxygen phase separation.  The grid covers the entire mass range
expected  for  carbon-oxygen  white  dwarfs,  from  0.52  to  $1.0  \,
M_{\sun}$, and it is based on a detailed and self-consistent treatment
of these energy  sources.  Except for the $1.0  \, M_{\sun}$ sequence,
the  history  of progenitor  stars  has  been  taken into  account  by
evolving initial stellar  configurations in the mass range  1 to $5 \,
M_{\sun}$ from the ZAMS all  the way through the thermally pulsing AGB
and mass loss phases. Because of the full calculation of the evolution
of progenitor stars, the  white dwarf sequences incorporates realistic
and consistent carbon-oxygen profiles --- of relevance for an accurate
computation of the energy  released by carbon-oxygen phase separation.
In addition,  detailed non-gray model  atmospheres are used  to derive
the outer boundary  condition for the evolving sequences.   At the low
luminosities  where  the process  of  $^{22}$Ne sedimentation  becomes
relevant, the outer boundary conditions influence the cooling times.

We find that $^{22}$Ne  sedimentation has notable consequences for the
cooling  times of  cool white  dwarfs  characterized by  a high  metal
content in their interiors. The related energy release strongly delays
their cooling. The precise value of  the delays depends on the mass of
the  white  dwarf, its  luminosity  and  on  the metal  content.   For
instance, because  of their larger gravities, the  impact of $^{22}$Ne
sedimentation  starts  earlier  in  more  massive  white  dwarfs.   In
particular, appreciable delays in  the cooling rates start to manifest
themselves  at  luminosities  of  $\log(L/L_{\sun}) \approx  -3.5$  to
$-4.2$.  In general, the magnitude  of the delays in the cooling rates
resulting from  $^{22}$Ne sedimentation is comparable  (or even larger
in the case of $Z=0.06$)  to the delays induced by carbon-oxygen phase
separation.   At the  approximate location  of the  faint peak  in the
white dwarf luminosity function of  NGC 6791, delays between 1 and 1.5
Gyr  are expected  as a  result of  $^{22}$Ne sedimentation  only.  As
recently shown in Garc\'ia--Berro et al (2010), the occurrence of this
process  in the  interior of  cool  white dwarfs  is a  key factor  in
solving the longstanding age discrepancy of NGC 6791.

In summary, we  find that the evolution of  cool white dwarfs stemming
from  progenitor  stars  with  super-solar  metallicity,  is  strongly
modified  by the  energy released  from $^{22}$Ne  sedimentation.  The
resulting delays in  cooling times of such white  dwarfs are important
and have to be taken  into account in age determinations of metal-rich
clusters from the cooling sequence  of their white dwarfs. The grid of
evolutionary  sequences  we  have  presented  here is  the  first  one
intented   for  such   a   purpose,  and   incorporates  the   effects
carbon-oxygen  phase  separation and  $^{22}$Ne  sedimentation in  the
evolution of these stars.

\acknowledgements

We acknowledge the comments and suggestions of our referee, which help
us to improve  the original version of this  paper.  This research was
supported  by  AGAUR,  by  MCINN  grants  AYA2008--04211--C02--01  and
AYA08-1839/ESP,  by  the  European  Union  FEDER  funds,  by  AGENCIA:
Programa de  Modernizaci\'on Tecnol\'ogica BID 1728/OC-AR,  and by PIP
2008-00940  from CONICET.  LGA also  acknowledges a  PIV grant  of the
AGAUR of the Generalitat de Catalunya.

\end{document}